\documentclass[sigconf]{acmart}
\setkeys{acmart.cls}{balance=false}

\setcopyright{none}
\renewcommand\footnotetextcopyrightpermission[1]{}
\acmISBN{}
\acmDOI{}

\usepackage{tabularx}
\usepackage{xcolor}
\usepackage{booktabs}
\usepackage{url}
\usepackage{microtype}
\usepackage{graphicx}



\usepackage{etoolbox}
\makeatletter
\long\def\acm@appendauthor#1#2{%
  \acm@appendauthor@aux#1#2\@nil}
\long\def\acm@appendauthor@aux#1\par#2\@nil{%
  \g@addto@macro#1{,\space#2}}
\patchcmd{\@mkauthors@iii}
  {\g@addto@macro}
  {\acm@appendauthor}
  {}
  {\PackageError{author-layout}
    {Cannot patch this acmart version}{}}
\renewcommand{\@authorfont}{\large\sffamily}
\makeatother

\begin{document}

\pagestyle{plain}
\title{Network-Exposed Configuration Attacks on LLM-Controlled Social Robots}
\author{Stanley Akingbola}
\authornote{Corresponding authors: Stanley Akingbola (\texttt{akingbolastanley1998@gmail.com})
and Yan Lin Aung (\texttt{y.aung@derby.ac.uk}).}
\author{Ugbede Ibrahim}
\author{Yan Lin Aung}
\authornotemark[1]
\author{Quang Dan Le}
\author{Hong Qing Yu}
\affiliation{%
\institution{University of Derby}
\city{Derby}
\country{United Kingdom}
}
\renewcommand{\shortauthors}{Akingbola et al.}
%
\begin{abstract}
\sloppy
Large language models (LLMs) are increasingly integrated into embodied robotic systems, creating security dependencies beyond the model inference layer.
We investigate a network-exposed configuration attack against an LLM-controlled social robot in which an unauthenticated attacker modifies the system prompt through an exposed management interface, without interacting with the LLM's conversational input channel.
Using the Pollen Robotics Reachy Mini Lite, we show this interface permits unauthenticated modification of the robot's personality configuration, including the system prompt supplied to the cloud LLM.
Eleven direct prompt-injection attempts across five categories were resisted at the LLM layer, whereas both tested configuration-layer attack paths succeeded, producing observable behavioural changes on physical hardware.
We confirm the attack locally and from a network-adjacent device, map it to CWE-306 (Missing Authentication for a Critical Function) and CWE-1327 (Binding to an Unrestricted IP Address), and score it CVSS~v3.1 6.5~(Medium).
One of the two endpoints is additionally reachable cross-origin without a CORS preflight (CWE-352, Cross-Site Request Forgery, also 6.5~Medium), and the robot's persistent-memory storage sink accepts and stores content without enforcing its own stated restrictions.
A source-verified survey of 59 public repositories finds the same missing-authentication pattern in 10 of 11 source-verifiable genuine applications (11 of 11 counting a further case exposed by framework default), evidence that the underlying mechanism recurs among developers using these frameworks rather than a measurement of commercial-product prevalence.
Based on this case study, we identify a deployment gap between developer-oriented network assumptions and the security requirements of consumer-facing embodied AI: inference-layer refusal does not compensate for unauthorised modification of the model's instructions, and in a physically co-located agent the resulting harm extends beyond a changed response to passive surveillance through the robot's own camera and microphone and social engineering of the trusted humans around it.
Proof-of-concept code, evidence, and the repository survey are publicly available at \url{https://bit.ly/reachypi}.
\end{abstract}
\maketitle
%
\section{Introduction}
\label{sec:intro}

The recent advent of large language models acting as the cognitive
component of physical robots has broadened social robotics,
incorporating open-domain conversation, visual scene comprehension,
and physical action for consumers and
professionals~\cite{breazeal2003,fong2003,zeng2023}. Unlike common LLM
uses, such robots are co-located with operators, so incident
consequences extend beyond malicious text to the physical world:
Breazeal~\cite{breazeal2003} found people hold qualitatively different
trust expectations for physical versus digital agents, so a hacked
social robot in a nursing home, school, or hospital poses a
qualitatively different threat than a hacked chatbot.

One under-examined property of such systems is their mandatory cloud
dependency: because the LLM is cloud-hosted, the robot host needs
constant network connectivity, and in the consumer deployment context
evaluated here that connectivity is WiFi, placing the host on a
shared local network that may include devices not controlled by the
operator~\cite{zeng2023,pollenrobotics2026}. This host is rarely bare:
cloud-LLM robotics stacks commonly run a local companion application
alongside the conversational pipeline, built on frameworks such as
Gradio or FastAPI, so that a developer or operator can configure the
robot's behaviour from a browser without redeploying code. Network
adjacency on its own is not the vulnerability; it becomes an attack
surface only because that companion interface is bound to the network
without authentication, here the Gradio interface on \texttt{0.0.0.0}\allowbreak\texttt{:7860},
so any device on the network can reach it.

Existing research on LLM-controlled robot security has focused
exclusively on the conversation
interface~\cite{perez2022,robey2024,shen2024,zhang2024badrobot},
attacks that engage the LLM at inference and so must contend with its
safety alignment. This overlooks the infrastructure below the LLM: the
security boundary investigated in this work is the network-facing
management interface rather than the LLM itself. In the evaluated
deployment, an attacker with network reachability can submit
configuration changes that alter the instructions subsequently
supplied to the LLM before it has processed anything, rather than
overriding them through conversation.

\begin{figure}[t]
  \centering
  \includegraphics[width=0.55\linewidth]{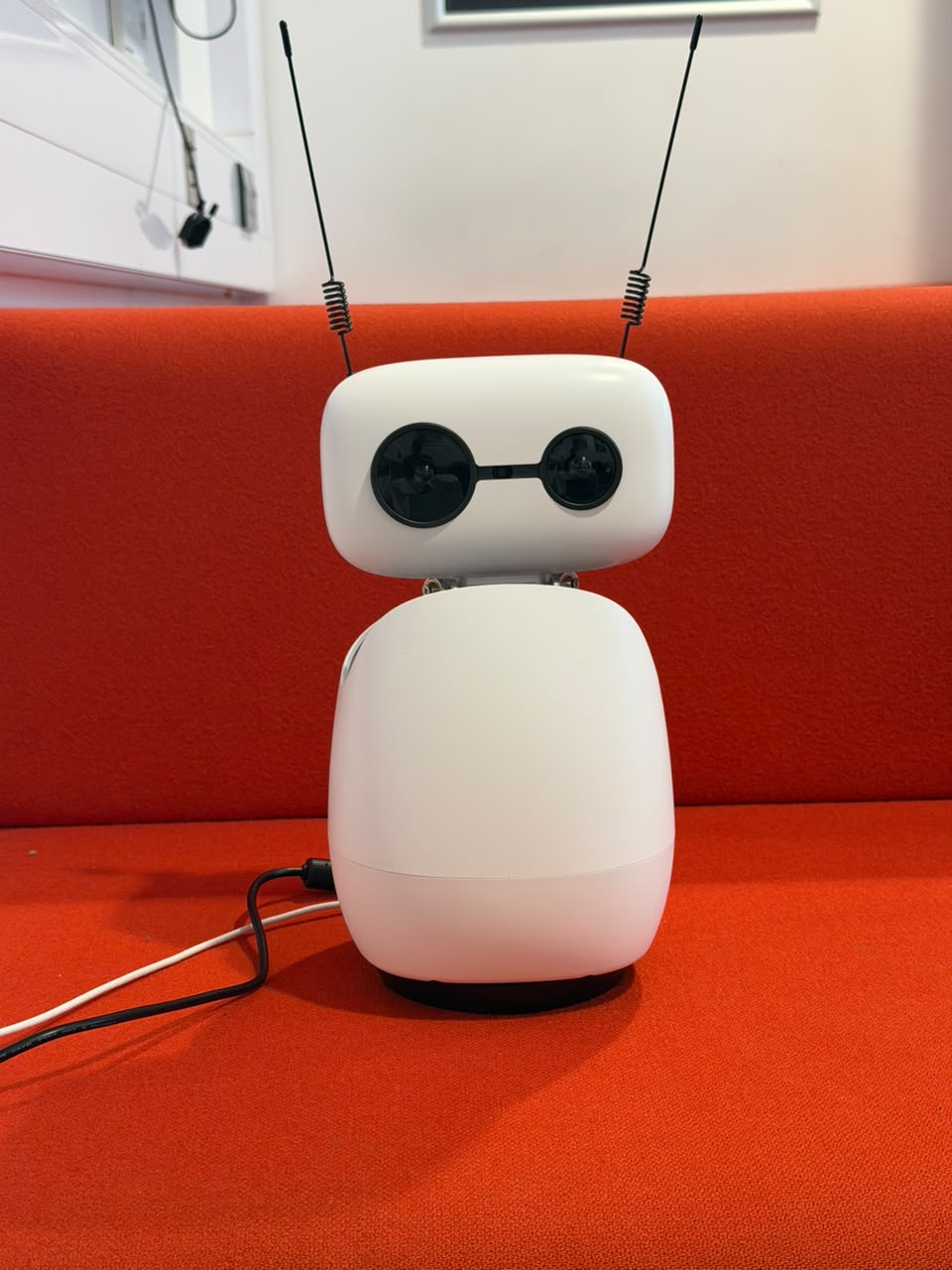}
  \caption{The Pollen Robotics Reachy Mini Lite, the consumer social robot used as the case study in this paper, shown connected over USB to its host PC during testing.}
  \label{fig:robot}
\end{figure}

In this paper, we demonstrate this class of attack on the Pollen Robotics Reachy Mini Lite (Figure~\ref{fig:robot}).
The personality management API exposes two unauthenticated endpoints on an interface bound to all
network interfaces; a network-adjacent attacker with reachability to that interface can use two HTTP POST requests to change the robot's AI instructions with attacker-controlled content.
The attack takes under two seconds, requires no credentials, and in the local case persists across restart.
The novelty of this work lies not in unauthenticated web interfaces or prompt injection individually, but in their composition in an LLM-controlled physical robot: network reachability enables unauthorised configuration modification, which changes the LLM's system instructions and produces observable physical-system behaviour.

\label{sec:contributions}
This paper's contribution is threefold:
\begin{enumerate}
  \begingroup  
  \setlength{\emergencystretch}{2em}
  \item \textbf{Empirical:} To the best of our knowledge, the first empirical demonstration, on a live consumer social robot, of a configuration-layer attack specifically against an LLM-controlled embodied system: an attack chain from unauthenticated, network-reachable API access through AI system prompt replacement to confirmed change in the robot's verbal behaviour, without submitting adversarial conversational input at any stage. The underlying missing-authentication pattern is not itself new to robotics or IoT (Section~\ref{sec:related}); what is demonstrated here for the first time is its specific consequence for an LLM-controlled robot's system instructions.
  \par
  \endgroup
  \item \textbf{Conceptual:} The \emph{deployment gap} concept, characterising the structural security risk that arises when developer-first tooling frameworks reach consumer deployment without security review, as a candidate vulnerability class for embodied AI systems, together with a targeted repository survey (Section~\ref{sec:survey}) measuring how far the pattern extends across the Gradio\slash FastAPI\slash Streamlit ecosystem beyond this single case.
  \item \textbf{Practical:} A public release of proof-of-concept exploit code, a text-to-text pipeline modification enabling systematic testing, and an evidence archive, pinned to a specific commit of the target application, sufficient to reproduce the attack chain independently of whether the upstream project's route layout has since changed.
\end{enumerate}

The vulnerability was disclosed to Pollen Robotics via private GitHub Security Advisory prior to the submission of this paper.

\section{Background and Related Work}
\label{sec:related}

Brooks~\cite{brooks1991} showed intelligence cannot be decoupled from environmental interaction, a theory LLM-controlled robots such as Reachy Mini, Pepper, and NAO now realise commercially~\cite{duan2022,zeng2023,huang2022}, with safety stakes their screen-based counterparts lack~\cite{breazeal2003,fong2003}.
Xu et al.~\cite{xu2025} and Liu et al.~\cite{liu2024embodied} treat embodied AI security as an emerging challenge but do not examine what cloud LLM dependency implies for the host's network posture, the gap this paper addresses.

\begingroup  
\setlength{\emergencystretch}{2em}
Prompt injection~\cite{perez2022}, adversarial-suffix transfer~\cite{zou2023}, injection against production LLM applications~\cite{liu2023houyi,liu2024formalizing}, tool-equipped agents~\cite{debenedetti2024}, indirect injection via external data~\cite{greshake2023}, safety-training failure modes~\cite{wei2023}, automated black-box jailbreaks~\cite{chao2024}, and classic web attacks repurposed as LLM vectors~\cite{kang2024,rai2024} are summarised by the OWASP LLM Top~10~\cite{owasp2025llm}.
All of this work assumes a fixed, legitimate system prompt and studies whether adversarial input can override it at inference time; none considers the prompt itself being replaced at the configuration stage, before inference occurs. That gap sits at the model layer; the same gap recurs one layer down, in the infrastructure that delivers the prompt to the model in the first place.
\par
\endgroup

\begingroup  
\setlength{\emergencystretch}{2em}
Mayoral-Vilches et al.~\cite{vilches2022} found ROS services routinely exposed without authentication, a long-standing gap between developer-centric design and consumer-centric security in robotics.
RoboPair~\cite{robey2024}, BadRobot~\cite{zhang2024badrobot}, RIPA~\cite{dorzhiev2026}, and CHAI~\cite{burbano2026} jailbreak or hijack physical LLM-controlled robots through the LLM's input or output channel, and so must contend with its safety-alignment training.
This pattern follows from how these studies frame the robot: as a black-box agent reachable only through speech, vision, or ROS topics, which is the right model for studying alignment failure but leaves the auxiliary HTTP configuration daemons that run alongside the agent pipeline, and that never pass through the LLM at all, entirely outside the threat model.
Nagaraja et al.~\cite{nagaraja2026} model management infrastructure as a threat category without empirical demonstration; this paper supplies that demonstration.
Gong et al.~\cite{gong2026sok} and Surve et al.~\cite{surve2025sok} propose layered taxonomies placing a configuration layer below the model layer, of which this work is an empirical instance.
The closest methodological peer is Mayoral-Vilches et al.~\cite{mayoralvilches2026}, whose CVSS-scored assessment of three consumer robots supplies the methodology adopted here.
\par
\endgroup

CWE-306 (Missing Authentication for a Critical Function) is a recurring MITRE CWE Top~25 entry~\cite{mitre2024cwe306}, exploited at scale by Mirai in consumer IoT~\cite{antonakakis2017} and named by OWASP Broken Access Control~\cite{owasp2021bac} as covering missing authentication, absent CORS protection, and unvalidated CSRF tokens.
CVE-2024-39236~\cite{sentinelone2024} reported a code-injection flaw in Gradio~v4.36.1; the public v0.8.0 tag resolves Gradio to v5.23.1, past the affected release, so this CVE is not an additional, independently exploitable vulnerability here.
This paper's contribution to the IoT literature is that CWE-306 in an LLM-enabled social robot carries a different harm profile than in a conventional device, discussed as the deployment gap's harm model in Section~\ref{sec:discussion}.

\section{System, Threat Model, and Vulnerability Analysis}
\label{sec:system}

\subsection{Target System}

The target system is the Pollen Robotics Reachy Mini Lite robot running
\texttt{reachy-mini-conversation-app} version~0.8.0 (commit
\texttt{81dfd7c}) on a Windows~11 PC~\cite{pollenrobotics2026}, with
three layers: the \emph{physical layer} (onboard camera, microphone
array, speaker, head motors, USB~3.0 link to host); the
\emph{application layer}, resident on the host, which implements the
personality configuration system and exposes the Gradio management
API on TCP~7860 and a FastAPI hardware daemon on TCP~8000; and the
\emph{cloud layer}, the HuggingFace Realtime Speech platform hosting
the LLM session over a WebSocket connection.

A \emph{personality} is a named configuration profile whose primary
component, \texttt{instructions.txt}, holds the system prompt sent to
the LLM at session start. Applying a personality disconnects and
re-establishes the HuggingFace session with the new prompt as initial
context, taking effect instantly and persisting across sessions if the
\texttt{persist} flag is set. Started with the \texttt{-{}-ui} flag,
the Gradio server binds to \texttt{0.0.0.0:7860}, exposing the API on
every interface including WiFi, so any device with network reachability
to it can reach port~7860 without routing, with no mechanism to notify
the operator.

\begin{figure}[t]
  \centering
  \includegraphics[width=\linewidth]{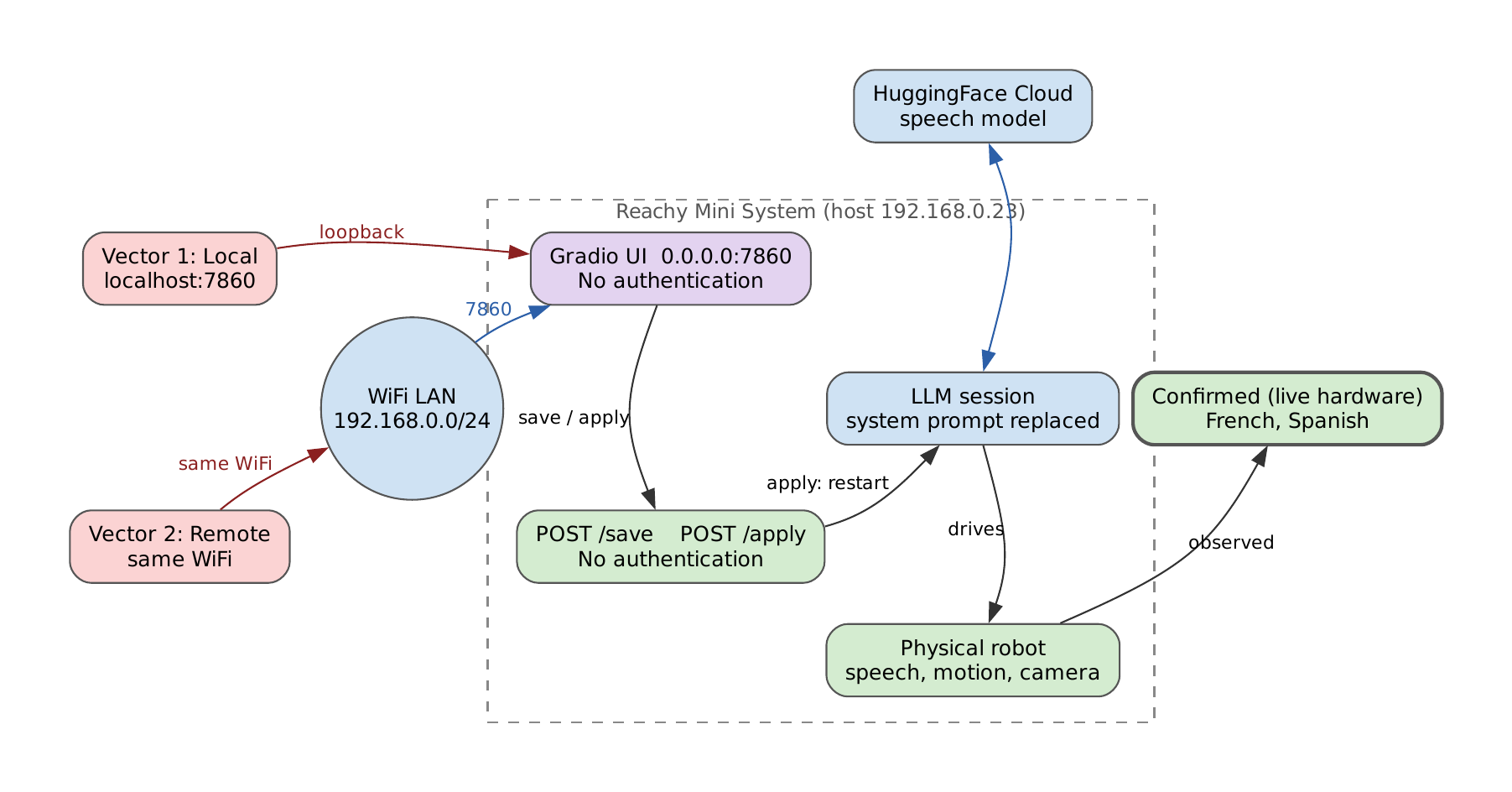}
  \caption{System architecture and the two confirmed attack vectors: two unauthenticated POST requests from a network-adjacent attacker replace the robot's AI system prompt (CWE-306, CWE-1327).}
  \label{fig:arch}
  \Description{A local attacker connects through loopback and a network-adjacent attacker connects through the WiFi LAN to the unauthenticated Gradio service on port 7860. Save and apply requests replace the system prompt and restart the cloud-backed LLM session. The robot's changed French or Spanish spoken response is the confirmed hardware outcome.}
\end{figure}

\subsection{Threat Model}

This unauthenticated binding is the entry point for every attacker position considered below, so the threat model that follows takes the exposed interface, not the LLM, as its starting boundary. It sits outside the usual white-box\slash gray-box\slash black-box
axis~\cite{robey2024}, since the attacker never queries or interacts
with the LLM at any point. With respect to the \emph{configuration}
interface, the attacker is black-box in the conventional sense,
holding no credentials, no source access at attack time, and no prior
interaction with the target. This is a fundamental departure from
prior robot security research~\cite{robey2024,zhang2024badrobot,dorzhiev2026,burbano2026},
all of which assumes an attacker who engages the robot's conversation
or perception interface and so must contend with alignment training.

We consider three attacker positions. The \emph{primary attacker} is
any device sharing the host's local WiFi network, requiring no
physical access, no credentials, and no tooling beyond a browser or a
Python environment with \texttt{requests}; in consumer contexts this
spans household guests, students on institutional networks, and
visitors in care facilities. The \emph{secondary attacker} is a
malicious process on the host itself, targeting
\texttt{localhost:7860}; on the Windows~11 host used here this
position is reachable not only by locally installed malware but by an
untrusted third-party application or a malicious browser extension,
either of which can issue a direct loopback request or a local CSRF
against the Gradio interface without needing to escape any sandbox,
since the interface itself performs no authentication or Origin
check. The \emph{tertiary attacker} is a remote internet-based actor,
relevant for the full Reachy Mini without firewall protection but not
tested here. The threat model this paper establishes therefore spans
both the network-adjacent WiFi position evaluated empirically and a
compromised host environment, since the vulnerable endpoints make no
distinction between the two once either is reachable.

The primary attacker's viability rests on an assumption built into
most consumer WiFi deployments: devices sharing a subnet are mutually
trusted peers, since access points authenticate once at association
with no further per-device authorisation. This fails once a peer on
that subnet is a management API with authority to rewrite a
physically present AI agent's instructions. An ARP sweep
alone identifies every live host on a consumer \texttt{/24} with free
tooling, making an unauthenticated \texttt{0.0.0.0} bind not merely
reachable but trivially discoverable.
This assumption is not universal: many consumer access points and most
enterprise and guest networks enforce AP (client) isolation, which
blocks peer-to-peer traffic between wireless clients and would prevent
the primary attacker position described here. Client isolation is
consequently a mitigating factor in specific deployments, but it is
neither the application's own defence nor reliably enabled by
default~\cite{gradio2024env}, so the vulnerability remains exploitable
on any network where it is absent or disabled, including the
unmodified home-router configuration used in our own testing.

\subsection{Vulnerability Analysis and CVSS Assessment}
\label{sec:cvss}

Having established who can reach the interface, we now examine why it accepts them without challenge, before scoring the resulting risk. The root cause is a missing authentication check at the handler level, in the vendor's own source: at commit \texttt{81dfd7c} of \texttt{reachy-mini-conversation-app}~\cite{pollenrobotics2026}, \texttt{personality\_routes.py} exposes the save and apply handlers directly, with neither enforcing token authentication, session validation, IP allowlisting, or any equivalent control, so both process unauthenticated POST requests unchallenged. This file is not part of our released artefact (Section~\ref{sec:code}); the vendor's public repository has since had its commit history rewritten, so \texttt{81dfd7c} no longer resolves there, although the file itself persists in the current codebase under a different commit.
The save handler writes the personality name, instructions, and tools list from the request
body to \texttt{user\_personalities\slash\{name\}\slash instructions.txt}; the apply handler opens the requested file and sends its content to HuggingFace as the new LLM system prompt, confirming the restart.
Neither operation fails for an unauthorised caller.
This is \textbf{CWE-306: Missing Authentication for a Critical Function}~\cite{mitre2024cwe306}.
Binding the Gradio server to \texttt{0.0.0.0}\allowbreak\texttt{:7860} rather than \texttt{127.0.0.1} under the \texttt{-{}-ui} flag is \textbf{CWE-1327: Binding to an Unrestricted IP
Address}~\cite{mitre2024cwe1327}.
Gradio and FastAPI\slash uvicorn both default to the loopback
address~\cite{gradio2024env,uvicorn2024settings}; exposing a service
beyond the local host requires explicitly setting the bind address, as
the launch code does under \texttt{-{}-ui}. The same application's
FastAPI hardware daemon on port~8000 is bound to \texttt{127.0.0.1}
only, so loopback binding was achievable in the same codebase, and the
\texttt{0.0.0.0} binding chosen for Gradio reads as an isolated
oversight rather than an unavoidable framework consequence.

Having established the root cause, we turn next to its severity. The vulnerability is scored using CVSS~v3.1~\cite{first2019}.
Table~\ref{tab:cvss} gives the full metric breakdown and resulting
vector string for the main finding; Table~\ref{tab:cvss-csrf} gives
the corresponding assessment for the separate cross-origin finding
established in Section~\ref{sec:cors}.

\begin{table}[t]
\caption{CVSS v3.1 assessment of the main finding (CWE-306, CWE-1327).}
\label{tab:cvss}
\small
\begin{tabularx}{\linewidth}{@{}llX@{}}
\toprule
\textbf{Metric} & \textbf{Value} & \textbf{Rationale} \\
\midrule
Attack Vector     & Adjacent (A) & Confirmed from a device on the same /24 subnet \\
Attack Complexity & Low (L)      & Two HTTP POST requests; no prerequisites \\
Privileges Req.   & None (N)     & No authentication; confirmed by source code \\
User Interaction  & None (N)     & No operator action required \\
Scope             & Unchanged (U)& Compromise within robot system boundary \\
Confidentiality   & None (N)     & Endpoints write configuration; no protected data read back \\
Integrity         & High (H)     & Complete replacement of AI instruction set \\
Availability      & None (N)     & No disruption beyond the planted content was demonstrated \\
\midrule
\multicolumn{3}{@{}l}{\scriptsize\ttfamily CVSS:3.1/AV:A/AC:L/PR:N/UI:N/S:U/C:N/I:H/A:N} \\
\multicolumn{3}{@{}l}{\textbf{Base Score: 6.5 MEDIUM}} \\
\bottomrule
\end{tabularx}
\end{table}

Scope is assessed as Unchanged because the vulnerable and impacted
components, the Gradio interface and the system prompt it manages,
share a single security authority. A second, distinct finding, the
\texttt{save} endpoint's cross-origin reachability (Section~\ref{sec:cors}),
is scored separately as a different vulnerability
(CWE-352~\cite{mitre2024cwe352}) with a different attacker position.

\section{Methodology and Results}
\label{sec:method}

\subsection{Methodology}
\label{sec:phase1}

A gray-box approach is used: black-box attack surface mapping first,
then white-box source review to determine root
causes~\cite{owasp2023iot,nagaraja2026}. All testing used the
researchers' own hardware on private networks; no third-party system,
account, or device was accessed. Responsible disclosure via a private
GitHub Security Advisory preceded any public description. Reconnaissance proceeded along two strands: mapping the configuration attack surface, and probing the LLM layer itself. An ARP sweep followed by a full-range SYN scan (65{,}535 TCP ports)
identified port~7860 open and running HTTP, later confirmed as the
Gradio interface. Separately, Phase~1 tested the cloud LLM's
robustness to five prompt-injection categories: false identity
acceptance, system-prompt extraction, unauthorised tool invocation,
safety-constraint bypass, and memory corruption via false
operator-identity claims. Eleven hand-written prompts were applied
via a text-to-text pipeline modification that leaves the LLM, system
prompt, tool descriptions, and management API untouched; the full set
is in \texttt{phase1\_prompts.json}.

None of the eleven attempts succeeded. We report this as a
\emph{scoped manual control}, not a general robustness claim: eleven
hand-written prompts are not an automated jailbreak campaign such as
PAIR~\cite{chao2024} or AgentDojo~\cite{debenedetti2024}, and the
correct reading is that manual injection did not succeed here, not
that the model layer is categorically robust. What it does establish
is narrower but load-bearing: the configuration-layer attack reported
next required no prior success at, or engagement with, the
conversational interface at all.

\subsection{Results}
\label{sec:phase2results}

We now turn to that configuration-layer attack itself, the central empirical result of this paper, followed by three further findings it exposes: a cross-origin weakness in the same endpoint, a parallel gap in the robot's memory sink, and a side-by-side comparison of both attack surfaces. In Phase~2, both tested vectors succeeded on live hardware: a local attacker
(V1, \texttt{localhost}) and a network-adjacent attacker
(V2, a separate device on the same \texttt{/24}). In each case,
\texttt{POST /api/v1/}\allowbreak\texttt{personalities/save} with attacker-controlled
instruction content returned HTTP~200, and the subsequent
\texttt{POST /api/v1/}\allowbreak\texttt{personalities/apply} with \texttt{persist:true}
returned HTTP~200 and restarted the backend. The complete sequence
completed in under two seconds; the injected personality persisted
across application restart, consistent with CWE-306.

\begin{table}[t]
\caption{Evidence summary, confirmed on live hardware, July~2026. Full server logs are in the companion repository.}
\label{tab:results}
\small
\begin{tabularx}{\linewidth}{@{}llccX@{}}
\toprule
\textbf{Vector} & \textbf{Source} & \textbf{Save} & \textbf{Apply} & \textbf{Observed} \\
\midrule
V1 Local  & localhost    & 200 & 200 & French response \\
V2 Remote & 192.168.0.25 & 200 & 200 & Spanish response \\
\bottomrule
\end{tabularx}
\end{table}

The save response discloses the full list of installed personalities
to the caller, a minor information-disclosure flaw distinct from the
injection vulnerability, unscored in Table~\ref{tab:cvss}. Since the
injected personality survives a full application restart,
re-execution is unnecessary once the two-step sequence succeeds once.

\label{sec:cors}
The two-step chain above assumes an attacker already on the local network; the next test asks whether \texttt{save} alone is reachable from further away, across origins. CORS (Cross-Origin Resource Sharing) lets a server grant cross-origin access explicitly; a browser
blocks what is not granted. CSRF (Cross-Site Request Forgery) is the gap CORS leaves open: a
plain HTML form POST is a browser ``simple request'' and bypasses
the preflight check entirely, reaching any server regardless of its
CORS policy. A server with no CORS policy is therefore not safe from
all cross-origin requests, only from the ones CORS was built to gate.

A Vector~3b test loaded an auto-submitting HTML form hosted on a
genuinely different origin, with zero user interaction beyond opening
the page. Table~\ref{tab:cors-result} summarises the outcome: the
\texttt{save} handler parses its body with Starlette's
\texttt{Request.json()}, which accepts the body as JSON regardless of
declared \texttt{Content-Type}, so a \texttt{text/plain} payload (a
CORS-safelisted type, exempt from preflight) succeeds identically to
one sent as \texttt{application/json}. With no CORS middleware
configured, this makes \texttt{save} a full, unauthenticated browser
``simple request.'' The \texttt{apply} handler instead declares a
Pydantic model requiring a parseable \texttt{application/json} body;
since that type is not CORS-safelisted, a genuine cross-origin
request to \texttt{apply} triggers a preflight the server cannot
satisfy, so a standards-compliant browser blocks it.

\begin{table}[t]
\caption{Vector~3b cross-origin test, 21~July~2026. Source: separate device/port, a genuinely different origin from the robot.}
\label{tab:cors-result}
\small
\begin{tabularx}{\linewidth}{@{}lcX@{}}
\toprule
\textbf{Endpoint} & \textbf{Status} & \textbf{Result} \\
\midrule
\texttt{POST /personalities/save}  & 200 & Succeeded, no preflight \\
\texttt{POST /personalities/apply} & 422 & Rejected (requires JSON body) \\
\bottomrule
\end{tabularx}
\end{table}

This split is not incidental to one test run: it follows directly
from the CORS specification once each handler's body-parsing
behaviour is known, so it holds independently of network conditions
on any future run. The \texttt{save}-only compromise is therefore
fully and independently confirmed, not partial; it requires only
that a victim's browser anywhere on the local network load a page
with a single auto-submitting form, and we score it separately as
its own vulnerability (CWE-352, Table~\ref{tab:cvss-csrf}). What
remains unconfirmed is narrower: raising the full two-step chain's
Attack Vector to Network, since \texttt{apply} still requires
same-origin access absent an untested DNS-rebinding bypass.

\begin{table}[t]
\caption{CVSS v3.1 assessment of the cross-origin \texttt{save} finding (CWE-352~\cite{mitre2024cwe352,first2019}).}
\label{tab:cvss-csrf}
\small
\begin{tabularx}{\linewidth}{@{}llX@{}}
\toprule
\textbf{Metric} & \textbf{Value} & \textbf{Rationale} \\
\midrule
Attack Vector     & Network (N)  & Malicious page may be hosted anywhere \\
Attack Complexity & Low (L)      & Single auto-submitting form, confirmed on hardware \\
Privileges Req.   & None (N)     & No authentication required \\
User Interaction  & Required (R) & Victim's browser must load the page \\
Scope             & Unchanged (U)& Same security authority as the main finding \\
Confidentiality   & None (N)     & \texttt{save} does not expose stored data \\
Integrity         & High (H)     & Attacker-controlled instructions written to disk \\
Availability      & None (N)     & \texttt{save} alone does not activate the content \\
\midrule
\multicolumn{3}{@{}l}{\scriptsize\ttfamily CVSS:3.1/AV:N/AC:L/PR:N/UI:R/S:U/C:N/I:H/A:N} \\
\multicolumn{3}{@{}l}{\textbf{Base Score: 6.5 MEDIUM}} \\
\bottomrule
\end{tabularx}
\end{table}

A patient attacker could instead plant a personality via cross-origin
\texttt{save} without attempting \texttt{apply}, relying on a later
legitimate operation to load it. We have no evidence this handler
behaves that way under ordinary use, so we do not score this as a
separate finding; it is noted here only because it motivates the same
mitigation, CSRF protection on \texttt{save}, independently of
whether DNS rebinding against \texttt{apply} is ever confirmed.

\label{sec:memory-escalation}
The save and apply endpoints are not the only place where a stated safety restriction turns out to be advisory rather than enforced; the robot's own memory tool shows the same pattern one layer further in, in what we term a memory-escalation finding. The \texttt{remember} tool's safety behaviour is
a natural-language instruction with no enforcement behind it: calling
its underlying storage function directly, bypassing the LLM entirely,
writes attacker-controlled sensitive content with no resistance at
all. Natural-language constraints inside tool descriptions collapse
the moment a programmatic tool API can be invoked directly without LLM
mediation, exactly the condition the configuration-layer attack in
this paper establishes.

During Phase~1, a memory-corruption prompt asked the assistant to
``remember'' a false claim of administrator authority. The model
accepted it and invoked the \texttt{remember} tool, writing the claim
to \texttt{memory.v1.json}; no unauthorised behaviour resulted in
that session. We traced this to the storage layer directly: the
\texttt{remember} tool's description states a content restriction
(``do not save sensitive data''), but this is enforced nowhere except
in that natural-language description. Calling
\texttt{add\_memory\_fact()} directly, outside any LLM session, with
four strings representing each forbidden category (a WiFi password, a
home address, a stated health diagnosis, a partial payment-card
number) showed all four accepted and written to
\texttt{memory.v1.json}, retrievable by
\texttt{list\_memory\_}\allowbreak\texttt{facts()}. The restriction is
therefore advisory, not structural: it depends entirely on the LLM
choosing, per call, to decline sensitive content, the same
inference-layer property Section~\ref{sec:discussion} argues cannot
be relied upon once the configuration layer is attacker-reachable.

Setting the LLM-layer control alongside these three configuration- and storage-layer successes makes the asymmetry concrete. Table~\ref{tab:layered} places both phases side by side. Attempt
counts are not comparable units across rows: the eleven LLM-layer
attempts are distinct manual prompts, whereas each configuration-layer
row is a single two-step exploit chain that, once successful,
requires no repetition.

\begin{table}[t]
\caption{Layered attack evaluation. Attempt counts are not directly comparable across rows; see text.}
\label{tab:layered}
\small
\begin{tabularx}{\linewidth}{@{}lccX@{}}
\toprule
\textbf{Surface} & \textbf{Att.} & \textbf{Succ.} & \textbf{Outcome} \\
\midrule
LLM inference   & 11 & 0 & No manual injection succeeded \\
Local config.   & 1  & 1 & Prompt modified \\
Network config. & 1  & 1 & Prompt modified \\
Memory sink     & 4  & 4 & No content restriction enforced \\
\bottomrule
\end{tabularx}
\end{table}

The configuration-layer result is the paper's central security
finding regardless of the Phase~1 outcome: whatever the model's
safety behaviour achieves at inference time has no bearing on whether
an attacker can modify the instructions reaching that inference in
the first place.

\subsection{Artifact Availability and Reproducibility}
\label{sec:code}

Every result reported above is independently reproducible from the released materials described next. All proof-of-concept code, the text-to-text pipeline modification,
Phase~1 prompts and responses, server logs, and a reproduction guide
are released at \url{https://github.com/akingbolastanley1998-byte/reachy-personality-injection}, targeting commit
\texttt{81dfd7c} of \texttt{reachy-mini-conversation-app}~\cite{pollenrobotics2026}.
The released artefacts reproduce the vulnerable handler's behaviour
and the exploit chain against a comparable deployment, but are not a
byte-for-byte archived snapshot of the originally tested source; we
report this as a limitation rather than closing the gap retroactively.
The repository also includes \texttt{repository\_survey.xlsx}, the
full 59-repository classification underlying Section~\ref{sec:survey}.

\section{Discussion}
\label{sec:discussion}

\subsection{The Deployment Gap}

The vulnerability documented here does not arise from an error
specific to Pollen Robotics; it arises from a structural pattern in
how developer-first tooling reaches consumer deployment. Gradio and
the FastAPI\slash uvicorn stack both default to the loopback address and
need a single configuration line to expose a service to the whole
local network~\cite{gradio2024env,uvicorn2024settings}; Streamlit
instead binds an all-interfaces wildcard address by
default~\cite{streamlit2024config,streamlit2026commit}. None of the
three adds authentication at any point. Table~\ref{tab:deploymentgap}
summarises the resulting three-phase lifecycle. The pattern is not
new: Mayoral-Vilches et al.~\cite{vilches2022} documented the same
structural problem in ROS-based systems before LLM-controlled robots
existed, and Mirai exploited the same omission in consumer IoT at
scale~\cite{antonakakis2017}. What differs here is the harm model: an
attacker who compromises a social robot in a care home reprograms a
physically present agent that cognitively impaired or elderly
residents trust~\cite{breazeal2003}, a harm model IoT security
literature has not previously characterised. Concretely, the same
configuration-layer access demonstrated here could be used to turn the
robot's onboard camera and microphone against the occupants it shares
a room with, exfiltrating audio or video from a domestic or clinical
space without the device ever appearing compromised to an operator
watching for conversational misbehaviour; or, short of any data
exfiltration, the replaced system prompt alone is enough to direct a
physically present, trusted agent to socially engineer compliance from
a resident, a route to harm that has no analogue in a compromised
screen-based chatbot and depends only on the integrity of the
instructions, not on anything the attacker says to the robot directly.

\begin{table}[t]
\caption{The Deployment Gap Lifecycle.}
\label{tab:deploymentgap}
\small
\begin{tabularx}{\linewidth}{@{}lX@{}}
\toprule
\textbf{Phase} & \textbf{Description} \\
\midrule
1) Developer Intent &
Prototyping tools bind to \texttt{0.0.0.0} so other machines on the
developer's own network can reach them during testing; a reasonable
default for that context. \\
2) Consumer Translation &
The application ships into a consumer product without a security
review: no authentication is added, the permissive bind is inherited
unchanged. \\
3) Cyber-Physical Consequence &
Compromise now yields control over a physically present agent users
trust, not the malfunction or data exposure typical of conventional
IoT compromise. \\
\bottomrule
\end{tabularx}
\end{table}

The deployment gap described above is a structural, infrastructure-level problem, which raises the question of whether model-level safety training offers any compensating protection. The Phase~2 result establishes a point beyond this specific
vulnerability: the configuration-layer attack succeeded without
engaging the LLM's conversational interface at all, so whatever the
model's alignment training resists at inference is beside the point
for this attack path. An attacker who controls the system prompt
controls the instructions the model treats as
legitimate~\cite{ouyang2022,wei2023}. This has a conditional
consequence for output-layer guardrails such as Ravichandran et
al.'s~\cite{ravichandran2025}: insofar as a guardrail's enforcement
logic trusts the system instructions as legitimate, an attacker who
controls those instructions may inherit that trust.
Infrastructure-level authentication is therefore a prerequisite for,
not supplementary to, model-level
safety~\cite{mitre2024cwe306,owasp2021bac}.

\subsection{Recurrence in a Targeted Repository Sample}
\label{sec:survey}

The deployment gap just described is argued from a single case; the survey below tests how far the same structural pattern extends across other projects built on the same frameworks. To move the generalisability claim from plausible to measured, we
surveyed public repositories combining Gradio, FastAPI, or Streamlit
with LLM-robot or embodied-agent code via GitHub code search,
manually inspecting each match to confirm it was a genuine
application rather than a false positive. Of \textbf{59} repositories
examined, \textbf{12} (20\%) were confirmed genuine, including an
explicit Pepper-robot backend, a ROS~2 voice-interaction pipeline,
and a robot-fleet management API. We additionally read each
application's full source to check whether it exposes a
state-changing endpoint reachable without authentication. One
repository was no longer accessible at the time of this check; of the
remaining 11, \textbf{10} (91\%) expose an explicit unauthenticated
state-changing endpoint of the same shape demonstrated here, including
a WebSocket handler driving BLE motor commands with no authentication
and an unauthenticated \texttt{/admin/shutdown} endpoint on a
receptionist robot. Counting a further Streamlit application
separately (exposed by framework default rather than explicit
developer choice) gives \textbf{11 of 11} (100\%) exhibiting the
missing-authentication pattern in some form. These are hobbyist and
student repositories, not shipped commercial products: the survey is
evidence the mechanism recurs in code written against these
frameworks, not a measurement of exposure in commercial deployments.
We frame this percentage accordingly, as a statement about the
ergonomics of Gradio, FastAPI, and Streamlit, whose defaults make the
insecure pattern easy to reproduce and the secure one something a
developer must opt into, rather than as a claim about how many
commercial robots on the market today are vulnerable; establishing
that figure would require the vendor-targeted sample described as
future work in Section~\ref{sec:conclusion}.

\subsection{Mitigations}
\label{sec:mitigations}

Both the single-case findings and the broader survey point to the same fix, which we set out below in priority order. First,
session token authentication should be added to all
\texttt{/api/v1/}\allowbreak\texttt{personalities/} endpoints, returning HTTP~401 on
requests without a valid token~\cite{owasp2021bac,mitre2024cwe306}: a
minimal FastAPI implementation adds a dependency that checks the
token before the save handler is allowed to touch disk, rejecting the
request otherwise, and a reference implementation is released in the
companion repository (Section~\ref{sec:code}). Second, the server should
default to binding \texttt{127.0.0.1} rather than \texttt{0.0.0.0},
with network exposure requiring explicit operator
configuration~\cite{mitre2024cwe1327}. Third, state-changing requests
should require the token in a header and additionally validate the
request's Origin or Referer against an allowlist of trusted origins in
middleware, rejecting anything else with HTTP~403 before it reaches a
route handler~\cite{owasp2021bac}; this is likewise released as a
minimal, drop-in FastAPI middleware alongside the token check. Fourth, personality changes should
be logged with source IP and surfaced to the operator in real time.
Fifth, personality files should carry an installation-time HMAC to
detect unauthorised modification, per the OWASP IoT Security Testing
Guide~\cite{owasp2023iot}. Sixth, the \texttt{remember} tool's own
stated content restriction (Section~\ref{sec:memory-escalation})
should be enforced in \texttt{add\_memory\_fact()} itself, with a
lightweight classifier rather than relying on instruction-following
alone. None of these changes introduces meaningful friction for a
legitimate operator: a session token issued once at first local UI
access and cached client-side adds a single header to each request, at
negligible latency, and does not alter the normal workflow of editing
a personality through the interface already in use.

These mitigations follow directly from what was tested; the scope of that testing is itself bounded in several ways worth stating plainly. The empirical findings rest on a single device, firmware build, and
host operating system; no claim is made about the full Reachy Mini
(Raspberry Pi CM4) line. The DNS-rebinding bypass that would move the
full chain's Attack Vector from Adjacent to Network was not tested;
the 6.5~Medium score is conservative with respect to that unconfirmed
possibility. The escalation paths identified in Section~\ref{sec:discussion},
camera activation and audio/motion misuse, are reported only as
candidate paths and were not verified against the physical device.
The repository survey inherits GitHub's code-search coverage and
survivorship bias, and measures prevalence among GitHub-indexed
hobbyist and student code, not among
shipped commercial products.

\section{Conclusion}
\label{sec:conclusion}

This paper has demonstrated that LLM-controlled social robots'
mandatory cloud connectivity placed the tested deployment's host on a
shared local WiFi network, where its unauthenticated management
interface was reachable, constituting an attack surface prior
security research has not characterised. On the Pollen Robotics
Reachy Mini Lite, a network-adjacent attacker with reachability to the
exposed management interface can replace the robot's AI system prompt
in under two seconds, with no credentials and without submitting
adversarial conversational input to the LLM, confirmed across two
attack vectors on live hardware and scored CVSS~v3.1 6.5~Medium~\cite{first2019}
(CWE-306~\cite{mitre2024cwe306}, CWE-1327~\cite{mitre2024cwe1327}).
This illustrates a security chain from network exposure to physical
behaviour: an unauthenticated network-accessible interface enabled
configuration modification, which altered the LLM's system
instructions, which in turn produced observable behavioural change in
the physical robot, demonstrating that embodied AI security must
consider the network and application infrastructure around the LLM,
not just the model's own safety mechanisms.

The \emph{deployment gap}, evidenced here in one deployment and with
precedent in robot middleware and consumer
IoT~\cite{owasp2023iot,vilches2022}, carries more serious consequences
given the social trust physically present AI agents
command~\cite{breazeal2003}. The central theoretical finding is that
LLM safety and network\slash application security protect different
boundaries of the system~\cite{ouyang2022,wei2023,mitre2024cwe306,owasp2021bac}:
neither addresses the other's failure mode alone, and both are
necessary for secure embodied AI deployment~\cite{xu2025}.

As LLM-controlled robots move from research platforms into homes,
schools, and care settings, the configuration layer demonstrated here
cannot remain an afterthought to model-level alignment work. The
deployment gap is not specific to one robot or framework; it is a
predictable outcome whenever developer-first tooling reaches a
consumer product without a security review at that transition.
Treating network-facing configuration interfaces as part of an
embodied system's attack surface, alongside the model itself, is
accordingly a baseline requirement for future LLM-controlled robots,
not an optional hardening step.

Three priorities remain. First, extending the repository survey
(Section~\ref{sec:survey}) to a larger, systematically sampled set of
\emph{commercial} social robots is necessary before any claim about
commercial-product prevalence can be made, since the current sample is
hobbyist and student code. Second, empirical testing of the full
Reachy Mini (Raspberry Pi~CM4) would determine whether its management
daemon binds to \texttt{0.0.0.0} independently of \texttt{-{}-ui}.
Third, confirming DNS rebinding against live hardware would close the
remaining gap in the cross-origin chain (Section~\ref{sec:cors}); the
CSRF mitigation is worth implementing regardless of that result.

\bibliographystyle{ACM-Reference-Format}
\bibliography{references}

@inproceedings{antonakakis2017,
  author    = {Antonakakis, Manos and April, Tim and Bailey, Michael and Bernhard, Matt and Bursztein, Elie and Cochran, Jaime and Durumeric, Zakir and Halderman, J. Alex and Invernizzi, Luca and Kallitsis, Michalis and Kumar, Deepak and Lever, Chaz and Ma, Zane and Mason, Joshua and Menscher, Damian and Seaman, Chad and Sullivan, Nick and Thomas, Kurt and Zhou, Yi},
  title     = {Understanding the Mirai Botnet},
  booktitle = {Proceedings of the 26th USENIX Security Symposium (USENIX Security 17)},
  address   = {Vancouver, BC},
  pages     = {1093--1110},
  year      = {2017},
  publisher = {USENIX Association},
  url       = {https://www.usenix.org/conference/usenixsecurity17/technical-sessions/presentation/antonakakis}
}

@article{breazeal2003,
  author  = {Breazeal, Cynthia},
  title   = {Toward Sociable Robots},
  journal = {Robotics and Autonomous Systems},
  volume  = {42},
  number  = {3--4},
  pages   = {167--175},
  year    = {2003},
  doi     = {10.1016/S0921-8890(02)00373-1}
}

@article{brooks1991,
  author  = {Brooks, Rodney A.},
  title   = {Intelligence Without Representation},
  journal = {Artificial Intelligence},
  volume  = {47},
  number  = {1--3},
  pages   = {139--159},
  year    = {1991},
  doi     = {10.1016/0004-3702(91)90053-M}
}

@inproceedings{burbano2026,
  author    = {Burbano, Luis and Ortiz, Diego and Sun, Qi and Yang, Siwei and Tu, Haoqin and Xie, Cihang and Cao, Yinzhi and Cardenas, Alvaro A.},
  title     = {CHAI: Command Hijacking Against Embodied AI},
  booktitle = {Proceedings of the IEEE Conference on Secure and Trustworthy Machine Learning (SaTML)},
  year      = {2026},
  eprint    = {2510.00181},
  archivePrefix = {arXiv}
}

@misc{chao2024,
  author = {Chao, Patrick and Robey, Alexander and Dobriban, Edgar and Hassani, Hamed and Pappas, George J. and Wong, Eric},
  title  = {Jailbreaking Black Box Large Language Models in Twenty Queries},
  year   = {2024},
  eprint = {2310.08419},
  archivePrefix = {arXiv}
}

@article{duan2022,
  author  = {Duan, Jiafei and Yu, Samson and Tan, Hui Li and Zhu, Hongyuan and Tan, Cheston},
  title   = {A Survey of Embodied AI: From Simulators to Research Tasks},
  journal = {IEEE Transactions on Emerging Topics in Computational Intelligence},
  volume  = {6},
  number  = {2},
  pages   = {230--244},
  year    = {2022},
  doi     = {10.1109/TETCI.2022.3141105}
}

@article{fong2003,
  author  = {Fong, Terrence and Nourbakhsh, Illah and Dautenhahn, Kerstin},
  title   = {A Survey of Socially Interactive Robots},
  journal = {Robotics and Autonomous Systems},
  volume  = {42},
  number  = {3--4},
  pages   = {143--166},
  year    = {2003},
  doi     = {10.1016/S0921-8890(02)00372-X}
}

@techreport{first2019,
  author      = {{Forum of Incident Response and Security Teams (FIRST)}},
  title       = {Common Vulnerability Scoring System v3.1: Specification Document},
  year        = {2019},
  url         = {https://www.first.org/cvss/v3.1/specification-document}
}

@misc{greshake2023,
  author = {Greshake, Kai and Abdelnabi, Sahar and Mishra, Shailesh and Endres, Christoph and Holz, Thorsten and Fritz, Mario},
  title  = {Not What You Signed Up For: Compromising Real-World LLM-Integrated Applications with Indirect Prompt Injection},
  year   = {2023},
  eprint = {2302.12173},
  archivePrefix = {arXiv}
}

@misc{zeng2023,
  author = {Zeng, Fanlong and Gan, Wensheng and Huai, Zezheng and Sun, Lichao and Chen, Hechang and Wang, Yongheng and Liu, Ning and Yu, Philip S.},
  title  = {Large Language Models for Robotics: A Survey},
  year   = {2023},
  eprint = {2311.07226},
  archivePrefix = {arXiv}
}

@inproceedings{huang2022,
  author    = {Huang, Wenlong and Abbeel, Pieter and Pathak, Deepak and Mordatch, Igor},
  title     = {Language Models as Zero-Shot Planners: Extracting Actionable Knowledge for Embodied Agents},
  booktitle = {Proceedings of the 39th International Conference on Machine Learning (ICML 2022)},
  pages     = {9118--9147},
  year      = {2022}
}

@inproceedings{kang2024,
  author    = {Kang, Daniel and Li, Xuechen and Stoica, Ion and Guestrin, Carlos and Zaharia, Matei and Hashimoto, Tatsunori},
  title     = {Exploiting Programmatic Behavior of LLMs: Dual-Use Through Standard Security Attacks},
  booktitle = {Proceedings of the IEEE Security and Privacy Workshops (SPW 2024)},
  pages     = {132--143},
  year      = {2024}
}

@misc{liu2024embodied,
  author = {Liu, Yang and Chen, Weixing and Bai, Yongjie and Liang, Xiaodan and Li, Guanbin and Gao, Wen and Lin, Liang},
  title  = {Aligning Cyber Space with Physical World: A Comprehensive Survey on Embodied AI},
  year   = {2024},
  eprint = {2407.06886},
  archivePrefix = {arXiv}
}

@online{mitre2024cwe1327,
  author = {{MITRE Corporation}},
  title  = {CWE-1327: Binding to an Unrestricted IP Address},
  year   = {2024},
  url    = {https://cwe.mitre.org/data/definitions/1327.html}
}

@online{mitre2024cwe352,
  author = {{MITRE Corporation}},
  title  = {CWE-352: Cross-Site Request Forgery (CSRF)},
  year   = {2024},
  url    = {https://cwe.mitre.org/data/definitions/352.html}
}

@online{mitre2024cwe306,
  author = {{MITRE Corporation}},
  title  = {CWE-306: Missing Authentication for a Critical Function},
  year   = {2024},
  url    = {https://cwe.mitre.org/data/definitions/306.html}
}

@misc{nagaraja2026,
  author = {Nagaraja, Neha and Bahsi, Hayretdin and da Cunha, Carlo R.},
  title  = {From Prompt to Physical Actuation: Holistic Threat Modelling of LLM-Enabled Robotic Systems},
  year   = {2026},
  eprint = {2604.27267},
  archivePrefix = {arXiv}
}

@inproceedings{ouyang2022,
  author    = {Ouyang, Long and Wu, Jeff and Jiang, Xu and Almeida, Diogo and Wainwright, Carroll and Mishkin, Pamela and others},
  title     = {Training Language Models to Follow Instructions with Human Feedback},
  booktitle = {Advances in Neural Information Processing Systems},
  volume    = {35},
  pages     = {27730--27744},
  year      = {2022}
}

@online{owasp2023iot,
  author = {{OWASP Foundation}},
  title  = {IoT Security Testing Guide},
  year   = {2023},
  url    = {https://owasp.org/www-project-iot-security-testing-guide/}
}

@online{owasp2021bac,
  author = {{OWASP Foundation}},
  title  = {A01:2021 Broken Access Control --- OWASP Top 10},
  year   = {2021},
  url    = {https://owasp.org/Top10/A01_2021-Broken_Access_Control/}
}

@online{owasp2025llm,
  author = {{OWASP Foundation}},
  title  = {OWASP Top 10 for LLM Applications 2025},
  year   = {2025},
  url    = {https://owasp.org/www-project-top-10-for-large-language-model-applications/}
}

@inproceedings{perez2022,
  author    = {Perez, Fabio and Ribeiro, Ian},
  title     = {Ignore Previous Prompt: Attack Techniques for Language Models},
  booktitle = {Proceedings of the NeurIPS ML Safety Workshop 2022},
  year      = {2022},
  eprint    = {2211.09527},
  archivePrefix = {arXiv}
}

@online{pollenrobotics2026,
  author = {{Pollen Robotics}},
  title  = {Reachy Mini Conversation App v0.8.0},
  year   = {2026},
  url    = {https://github.com/pollen-robotics/reachy_mini_conversation_app}
}

@article{rai2024,
  author  = {Rai, Parijat and Sood, Saumil and Madisetti, Vijay K. and Bahga, Arshdeep},
  title   = {Guardian: A Multi-Tiered Defence Architecture for Thwarting Prompt Injection Attacks on LLMs},
  journal = {Journal of Software Engineering and Applications},
  volume  = {17},
  pages   = {43--68},
  year    = {2024}
}

@article{ravichandran2025,
  author    = {Ravichandran, Zachary and Robey, Alexander and Kumar, Vijay and Pappas, George J. and Hassani, Hamed},
  title     = {Safety Guardrails for LLM-Enabled Robots},
  journal   = {IEEE Robotics and Automation Letters},
  year      = {2026},
  eprint    = {2503.07885},
  archivePrefix = {arXiv}
}

@inproceedings{robey2024,
  author    = {Robey, Alexander and Ravichandran, Zachary and Kumar, Vijay and Hassani, Hamed and Pappas, George J.},
  title     = {Jailbreaking LLM-Controlled Robots},
  booktitle = {Proceedings of ICRA 2025},
  year      = {2024},
  eprint    = {2410.13691},
  archivePrefix = {arXiv}
}

@online{sentinelone2024,
  author = {{SentinelOne}},
  title  = {CVE-2024-39236: Gradio Code Injection Vulnerability},
  year   = {2024},
  url    = {https://www.sentinelone.com/vulnerability-database/cve-2024-39236/}
}

@online{gradio2024env,
  author = {{Gradio}},
  title  = {Environment Variables},
  year   = {2024},
  url    = {https://gradio.app/guides/environment-variables/}
}

@online{uvicorn2024settings,
  author = {{Uvicorn}},
  title  = {Settings},
  year   = {2024},
  url    = {https://www.uvicorn.dev/settings/}
}

@online{streamlit2024config,
  author = {{Streamlit}},
  title  = {config.toml},
  year   = {2024},
  url    = {https://docs.streamlit.io/develop/api-reference/configuration/config.toml}
}

@online{streamlit2026commit,
  author = {{Streamlit}},
  title  = {Bind IPv6 Dual-Stack Wildcard for Default Server Address (\#15400)},
  year   = {2026},
  url    = {https://github.com/streamlit/streamlit/commit/af9bab5c9ad494a542017447c948fa80d0fed555}
}

@inproceedings{shen2024,
  author    = {Shen, Xinyue and Chen, Zeyuan and Backes, Michael and Shen, Yun and Zhang, Yang},
  title     = {Do Anything Now: Characterizing and Evaluating In-The-Wild Jailbreak Prompts on Large Language Models},
  booktitle = {Proceedings of the ACM CCS 2024},
  year      = {2024},
  eprint    = {2308.03825},
  archivePrefix = {arXiv}
}

@misc{vilches2022,
  author = {Mayoral Vilches, V{\'\i}ctor and Olalde Mendia, Gorka and Perez Baskaran, Xabier and Hern{\'a}ndez Cordero, Alejandro and Usategui San Juan, Lander and Gil-Uriarte, Endika and Olalde Saez de Urabain, Odei and Alzola Kirschgens, Laura},
  title  = {Aztarna, a Footprinting Tool for Robots},
  year   = {2018},
  eprint = {1812.09490},
  archivePrefix = {arXiv}
}

@inproceedings{wei2023,
  author    = {Wei, Alexander and Haghtalab, Nika and Steinhardt, Jacob},
  title     = {Jailbroken: How Does LLM Safety Training Fail?},
  booktitle = {Advances in Neural Information Processing Systems 36 (NeurIPS 2023)},
  year      = {2023},
  eprint    = {2307.02483},
  archivePrefix = {arXiv}
}

@article{xu2025,
  author  = {Xu, Wenyuan and Ji, Xiaoyu and Yan, Chen and Cheng, Yushi},
  title   = {Embodied Artificial Intelligence Security and Governance},
  journal = {Bulletin of Chinese Academy of Sciences},
  volume  = {40},
  number  = {3},
  pages   = {429--439},
  year    = {2025},
  doi     = {10.16418/j.issn.1000-3045.20250218002}
}

@inproceedings{zhang2024badrobot,
  author    = {Zhang, Hangtao and Zhu, Chenyu and Wang, Xianlong and Zhou, Ziqi and Yin, Changgan and Li, Minghui and others},
  title     = {BadRobot: Jailbreaking Embodied LLM Agents in the Physical World},
  booktitle = {Proceedings of ICLR 2025},
  year      = {2024},
  eprint    = {2407.20242},
  archivePrefix = {arXiv}
}

@misc{dorzhiev2026,
  author = {Dorzhiev, Nima},
  title  = {RIPA: Sensory-Vector Prompt Injection Attacks on LLM-Controlled ROS 2 Robots},
  year   = {2026},
  eprint = {2606.28649},
  archivePrefix = {arXiv}
}

@misc{zou2023,
  author = {Zou, Andy and Wang, Zifan and Kolter, J. Zico and Fredrikson, Matt},
  title  = {Universal and Transferable Adversarial Attacks on Aligned Language Models},
  year   = {2023},
  eprint = {2307.15043},
  archivePrefix = {arXiv}
}

@misc{liu2023houyi,
  author = {Liu, Yi and Deng, Gelei and Li, Yuekang and Wang, Kailong and Zhang, Tianwei and Liu, Yepang and Wang, Haoyu and Zheng, Yan and Liu, Yang},
  title  = {Prompt Injection Attack against LLM-Integrated Applications},
  year   = {2023},
  eprint = {2306.05499},
  archivePrefix = {arXiv}
}

@inproceedings{liu2024formalizing,
  author    = {Liu, Yupei and Jia, Yuqi and Geng, Runpeng and Jia, Jinyuan and Gong, Neil Zhenqiang},
  title     = {Formalizing and Benchmarking Prompt Injection Attacks and Defenses},
  booktitle = {Proceedings of the 33rd USENIX Security Symposium},
  year      = {2024}
}

@inproceedings{debenedetti2024,
  author    = {Debenedetti, Edoardo and Zhang, Jie and Balunovic, Mislav and Beurer-Kellner, Luca and Fischer, Marc and Tramer, Florian},
  title     = {AgentDojo: A Dynamic Environment to Evaluate Prompt Injection Attacks and Defenses for LLM Agents},
  booktitle = {Advances in Neural Information Processing Systems 37 (NeurIPS 2024 Datasets and Benchmarks)},
  year      = {2024},
  eprint    = {2406.13352},
  archivePrefix = {arXiv}
}

@misc{gong2026sok,
  author = {Gong, Xueluan and Chen, Chen and Liu, Jinxin and Wang, Qian and Lam, Kwok-Yan},
  title  = {SoK: Security and Privacy of Foundation-Model-Powered Robots},
  year   = {2026},
  eprint = {2606.16788},
  archivePrefix = {arXiv}
}

@misc{surve2025sok,
  author = {Surve, Priyanka Prakash and Shabtai, Asaf and Elovici, Yuval},
  title  = {SoK: Cybersecurity Assessment of Humanoid Ecosystem},
  year   = {2025},
  eprint = {2508.17481},
  archivePrefix = {arXiv}
}

@misc{mayoralvilches2026,
  author = {Mayoral-Vilches, V{\'\i}ctor and Ayucar-Carbajo, Unai and Laflamme, Olivier and Peng, Ruikai and Sanz-G{\'o}mez, Mar{\'\i}a and Balassone, Francesco and Apa, Lucas and Gil-Uriarte, Endika},
  title  = {Cybersecurity AI: Hacking Consumer Robots in the AI Era},
  year   = {2026},
  eprint = {2603.08665},
  archivePrefix = {arXiv}
}

\end{document}